\documentclass[preprint]{elsarticle}
\newcommand{\R}{{\mathbb{R}}}
\newcommand{\N}{{\mathbb{N}}}
\usepackage{epsfig,epsf,amsfonts,amscd}

\begin{document}
\title{Boundary conditions for the states with resonant tunnelling 
across the $\delta'$-potential }
\author{
A.V. Zolotaryuk}
\address{
Bogolyubov Institute for Theoretical Physics,
National Academy of Sciences of Ukraine,
03680 Kyiv, Ukraine 
}
\date{\today}

\begin{abstract}
The one-dimensional Schr\"{o}dinger equation with the point potential in the form of the derivative of Dirac's delta function, 
$\lambda \delta'(x)$ with $\lambda$ being a coupling constant, is investigated. This equation is known to require an extension to the space of wave functions $\psi(x)$ discontinuous at the origin 
 under the two-sided (at $x=\pm 0$) boundary conditions given through the transfer matrix $\left(\begin{array}{cc} 
\! \! \!\! \! {\cal A}~~~0 \\
 0 ~~~{\cal A}^{-1}\! \!\end{array} \right)$ where 
${\cal A} = { 2+\lambda \over 2-\lambda }$.  However, the recent studies, where a resonant non-zero transmission across this potential has been established to occur on discrete sets 
$\{ \lambda_n \}_{n=1}^\infty$ in the $\lambda$-space, contradict to  these boundary conditions used widely by many authors. The present communication aims at solving this discrepancy using a more general form of boundary conditions. 
 \end{abstract}

\maketitle

\section{Introduction}

The Schr\"{o}dinger operators with singular
zero-range potentials attract a considerable interest beginning from the pioneering work of Berezin and Faddeev \cite{bf}. These operators (for details and references see book 
\cite{a-h}) describe point or contact interactions which are widely used in various applications to quantum physics 
\cite{do,ll,se1,ppz,kun}. Intuitively, these interactions are understood as sharply localized potentials, exhibiting a number
of interesting and intriguing features. Applications of these models
to condensed matter physics (see, e.g., \cite{ael,ex,es,ces})
 are of particular interest nowadays, mainly
 because of the rapid progress in fabricating nanoscale quantum devices.

In this paper we consider the one-dimensional Schr\"{o}dinger equation 
\begin{equation}
-\psi''(x) + V(x)\psi(x) =E \psi(x),
\label{1}
\end{equation}
where the prime stands for the differentiation with respect
to the spatial coordinate $x$ and $\psi(x)$ is the wave function
for a  particle of mass $m$ and energy $ E $ (we use units in which 
$\hbar^2/2m =1$). The point potential $V(x)$ has the form of the derivative of Dirac's delta function, i.e.,
\begin{equation}
V(x) = \lambda \delta'(x),~~ \delta'(x) \doteq d\delta(x)/dx,
\label{2}
\end{equation} 
where $\lambda$ is a coupling constant. 

Until recently there was a consensus that potential (\ref{2}) does not allow any transmission reflecting an incident quantum particle at all energies \cite{se2,pa,cnp}. In this case the particle states on the left $(\R^-)$ and the right $(\R^+)$ half-lines are called {\it separated}. However, recently in a series of papers 
\cite{c-g,zci,tn,zo,gm} it was established the existence of discrete resonance sets 
 in the $\lambda$-space at which the transmission across the barrier 
$V(x)$ becomes non-zero resulting in the existence of {\it non-separated} states. More precisely, if the distribution $\delta'(x)$ 
 is appropriately regularized by a sequence of finite-range functions
$\Delta'_\varepsilon(x) $ where $\varepsilon$ is a squeezing parameter, i.e., $\Delta'_\varepsilon(x) \to \delta'(x)$
in the sense of distributions, then in the zero-range limit 
(as $\varepsilon \to 0$)
Eq.~(\ref{1}) with potential (\ref{2}) admits a countable set
of resonances $\{\lambda_n \}_{n=1}^\infty$ with a partial transparency (see, e.g., Fig.~4 in \cite{c-g} where the peaks in the transmission across potential (\ref{2}) and its regularized versions 
$\Delta'_\varepsilon(x)$ are clearly depicted). Outside this set potential (\ref{2}) is opaque acting as a perfect wall. 
Moreover, as shown for some particular cases of regularizing sequences $\Delta'_\varepsilon(x) $ \cite{zci,tn} and proved rigorously in a general case \cite{gm} under the regularization
\begin{equation}
\delta'(x) = \lim_{\varepsilon \to 0}
\Delta'_\varepsilon (x) = \lim_{\varepsilon \to 0}\varepsilon^{-2} 
v (x / \varepsilon) 
\label{3}
\end{equation}
with a compactly supported profile $v(\xi)$, $\xi \in \R$,
satisfying the dipole-like properties:
\begin{equation}
\int_\R v (\xi) d\xi =0~~~\mbox{and}~~~
\int_\R  \xi v (\xi) d\xi =-1,
\label{4}
\end{equation}
 the structure of the resonance set $\{\lambda_n \}_{n=1}^\infty$ depends on the form of the function $v(\xi)$.
However, the existence of resonance sets in the transmission across the $\delta'(x)$ potential proved by using regularization (\ref{3}) conflicts with the widely cited result of \u{S}eba (see Theorem 4 in \cite{se2}), the proof of which has been revised very recently  by Golovaty and Hryniv \cite{gh}. As a result, these authors have proved that the $\delta'(x)$ potential defined through regularization 
(\ref{3}) is not necessary opaque, so that a non-zero transmission across potential (\ref{2}) indeed can occur. 
Nevertheless, the study of scattering properties of Eq.~(\ref{1}) with this potential using the distributional limit
\begin{equation}
\delta'(x) = \lim_{\varepsilon \to 0}
\Delta'_\varepsilon (x) = \lim_{\varepsilon \to 0}
{\delta(x + \varepsilon)- \delta(x - \varepsilon) \over 2
\varepsilon } 
\label{5}
\end{equation}
 demonstrates without doubt the absence of any transmission 
\cite{pa}.

\section{A rectangular model with two squeezing parameters}

The reason why in some cases the 
$\delta'(x)$ barrier has been treated as a fully reflecting wall and in other cases it was proved to be resonantly transparent can be explained if we introduce two squeezing parameters as follows. Let us construct the regularizing sequence consisting of a rectangular barrier and a rectangular well with width $l$ and height $l^{-2}$ separated by distance $\rho$. More precisely, we define the profile of this sequence as
\begin{equation}
\Delta'_{l\rho}(x) = {1 \over l(l+\rho) }\left[ 
u \left( {x  \over l } \right) 
- u \left({ x -\rho \over l} -1 \right) \right] ,
\label{6}
\end{equation} 
where $u(\xi) =1 $ if $ \xi \in (0,1)$ and $u(\xi) =0 $ otherwise. 
Here $l$ and $\rho$ serve as two independent squeezing parameters. Particularly, both the repeated limits of profile (\ref{6}) give the same $\delta'(x)$ function, i.e., 
\begin{equation}
\lim_{\rho \to 0} \lim_{l \to 0}\Delta'_{l \rho}(x) =
\lim_{l \to 0} \lim_{\rho \to 0}\Delta'_{l \rho}(x) =\delta'(x) .
\label{7}
\end{equation} 
However, the first limit illustrated by path 1 in Fig.~1 results in the complete reflection of an incident particle from the $\delta'(x)$ barrier, while the second limit shown in this figure by path 2 leads to the existence of a discrete resonance set 
$\{ \lambda_n \}_{n=1}^\infty$ in the $\lambda$-space, where the 
 transparency is non-zero \cite{c-g}. Surprisingly, both the results are correct and this riddle can be solved if we note that the wave function $\psi(x)$ in Eq.~(\ref{1}) must be {\it discontinuous} at the origin ($x =0$) and the {\it existence of resonance sets} is the result of {\it cancellation of divergences} coming from the kinetic energy operator $-d^2/dx^2$ and the singular potential term 
$\lambda \delta'(x) \psi(x)$. As a result, for each 
$\lambda =\lambda_n$, $n \in \N$, the limiting total Hamiltonian is no more the sum of the kinetic and potential terms. As shown below in detail,
there is no cancellation of divergences if we first squeeze the barrier and the well getting the $\delta$-functions separated by distance $\rho$ and then accomplish limit (\ref{5}), i.e., follow path 1 as illustrated in Fig.~1. Contrary, when
 we first bring together the barrier and the well, squeezing afterwords their width, i.e., follow path 2, we obtain the resonant tunnelling as a result of cancellation of divergences.  Consequently, starting from the same initial regularizing profile given by the pair of parameter values $l=1$ and 
$\rho =c$ where $c >0$ is arbitrary, one can obtain quite different results in dependence what path, i.e., regularizing sequence, has been chosen, either 1 or 2. Clearly, as shown in Fig.~1, the $\delta'(x)$ function can be obtained from this initial profile by many other ways, like paths 3, 4 or 5. One can expect that any sufficiently fast squeezing of the distance 
$\rho$ compared to squeezing the width $l$, as illustrated in Fig.~1 by path 4, will also result in the existence of resonance sets. 
Contrary, when the barrier-well width $l$ is squeezing faster than the distance $\rho$, like path 5 in Fig.~1, no resonances occur and the 
$\delta'(x)$ is completely opaque.

The situation with the existence or non-existence of resonance sets described above can be clarified if we consider an explicit solution of Eq.~(\ref{1}) with the finite-range potential 
$V_{l \rho}(x) \doteq \lambda \Delta'_{l\rho}(x)$ given by (\ref{6})
and analyze its zero-range limit (as $l \to 0$ and $\rho \to 0$). To this end we look for a positive-energy  solution
 of Eq.~(\ref{1}) with the potential $V_{l \rho}(x)$ in the form
\begin{equation}
\psi (x) =  \left\{ \begin{array}{lllll}
  {\rm e}^{{\rm i}k x} +R \, {\rm e}^{-{\rm i}kx}
 &&  \mbox{for}~~~ - \infty < x < 0 , \\
A_1 \, {\rm e}^{px} +B_1 \, {\rm e}^{-px}  && 
  \mbox{for}~~~~~ 0 < x <  l ,  \\
A_2 \, \sin(kx) +B_2 \, \cos(kx) && \mbox{for} ~~~~~ l<x<l+\rho ,\\
A_3 \,  \sin(qx)  + B_3 \, \cos(qx)  && 
\mbox{for}~~~~~ l+\rho < x < 2l+\rho , \\
T \, {\rm e}^{{\rm i} k x}  && \mbox{for}~~~~~
  2l+\rho  < x < \infty ,
\end{array} \right.
\label{8}
\end{equation}
where $R$ and $T$ are defined as the reflection  
and transmission coefficients (from the left), respectively, and 
\begin{equation}
  k \doteq \sqrt{E}~,~~ p \doteq \sqrt{\lambda l^{-2} -E}~ ,~~ 
q \doteq \sqrt{\lambda l^{-2} +E}~.
\label{9}
\end{equation}
The unknown coefficients $A_j$ and $B_j$, $j=1,\, 2, \, 3 $   can
be eliminated in a standard way by matching the solutions at the 
boundaries $x = 0 ,~ l, ~l+\rho,~ 2l+\rho   $. As a result, the 
solution of Eq.~(\ref{1}) can be written through the transfer matrix
$\Lambda$ connecting the boundary conditions for the wave function 
$\psi(x)$ and its derivative $\psi'(x)$ at $x=0$ and $x=x_0 \doteq
2l+\rho$: 
\begin{eqnarray}
\left( \begin{array}{cc} \psi(x_0)  \\
\psi'(x_0) \end{array} \right) 
 = \Lambda \left(
\begin{array}{cc} \psi(0)   \\
\psi'(0)   \end{array} \right), ~~~ \Lambda =  \left(
\begin{array}{cc} \Lambda_{11}~~ 
\Lambda_{12} \\
\Lambda_{21} ~~ \Lambda_{22} \end{array} \right) .
\label{10}
\end{eqnarray}
Here the matrix elements $\Lambda_{ij}$, $i,\, j =1,\, 2$, 
are given by
\begin{figure}
\centerline{\includegraphics[width=0.75\textwidth]{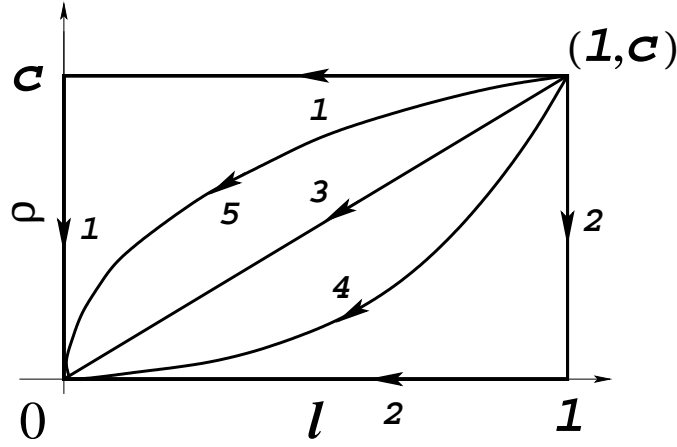}}
\vspace{2pt}
\caption{Schematics of different ways of regularization of the 
$\delta'(x)$ function shown by five paths 1, $\dots $ , 5 starting from the same rectangular profile given by parameter values $l=1$ and 
$\rho =c$
with any $c >0$. }
\label{fig1}
\end{figure}
\begin{eqnarray}
\Lambda_{11} &=& \left[
\cosh(pl) \cos(ql) + {p \over q} \sinh(pl) \sin(ql)\right]
\cos(k\rho) \nonumber \\
&+& \left[ {p \over k} \sinh(pl) \cos(ql) - {k \over q}
\cosh(pl) \sin(ql) \right]\sin(k\rho), \nonumber \\
\Lambda_{12} &=& \left[
{1 \over p} \sinh(pl) \cos(ql) + {1 \over q} 
\cosh(pl) \sin(ql) \right]\cos(k\rho) \nonumber \\
&+& \left[ {1 \over k} \cosh(pl) \cos(ql) - {k \over pq}
\sinh(pl) \sin(ql) \right]\sin(k\rho), \nonumber \\
\Lambda_{21} &=& \left[p \sinh(pl) \cos(ql) - q \cosh(pl) 
\sin(ql) \right]\cos(k\rho) \nonumber \\
&-& \left[ k \cosh(pl) \cos(ql) + {pq \over k} \sinh(pl) \sin(ql) \right]\sin(k\rho) , \nonumber \\
\Lambda_{22} &=& \left[
\cosh(pl) \cos(ql) - {q \over p} \sinh(pl) \sin(ql)\right]
\cos(k\rho) \nonumber \\
&-& \left[ {k \over p} \sinh(pl) \cos(ql) + {q \over k}
\cosh(pl) \sin(ql) \right]\sin(k\rho) ,
\label{11}
\end{eqnarray}
satisfying the condition 
\begin{equation}
\Lambda_{11}\Lambda_{22} - \Lambda_{12}\Lambda_{21} =1.
\label{12}
\end{equation}
Using the definition for the reflection and transmission 
coefficients given by Eqs.~(\ref{8}), one can rewrite 
Eq.~(\ref{10}) in the form
\begin{eqnarray}
 \left( \begin{array}{cc} T \\
 {\rm i}k T \end{array} \right)  {\rm e}^{{\rm i}kx_0}
 = \left( \begin{array}{cc} \Lambda_{11}~~ \Lambda_{12} \\
\Lambda_{21} ~~ \Lambda_{22}\end{array} \right) \left(
\begin{array}{cc} 1+R  \\
 {\rm i}k (1-R)  \end{array} \right) .
\label{13}
\end{eqnarray}
Solving next this matrix equation with respect to the coefficients 
$R$ and $T$, one finds their representation in terms of the matrix elements $\Lambda_{ij}$: 
\begin{equation}
R  =  -  \, {\Lambda_{11} -  \Lambda_{22} + {\rm i} 
(k \Lambda_{12}  + k^{-1} \Lambda_{21}) \over \Delta}
 ~~~\mbox{and}~~~
T ={ 2 \over \Delta } {\rm e}^{- {\rm i} k x_0 } 
\label{14}
\end{equation}
where $\Delta \doteq  \Lambda_{11} +  \Lambda_{22}
 - {\rm i} ( k \Lambda_{12}  - k^{-1} \Lambda_{21} )$.
Using here Eq.~(\ref{12}), one can easily check the validity of the conservation law $|R|^2 + |T|^2 =1$.

As follows from the form of expressions (\ref{11}), the most singular matrix element in the limit when $l \to 0$ and 
$\rho \to 0$ is $\Lambda_{21}$.  Therefore the analysis of this 
 limit should be started from this element. Consider first the second repeated limit (\ref{7}) illustrated in Fig.~1 by path 2. In this case
first $\sin(k\rho) \to 0$ and therefore the expression with the second square brackets in $\Lambda_{21}$ vanishes prior to the $l \to 0$ limit, whereas the terms in the first square brackets cancel out under the equation 
\begin{equation}
\tanh\sigma  =\tan\sigma ,~~\sigma \doteq \sqrt{\lambda}\, , 
\label{15}
\end{equation}
 found earlier in another way \cite{c-g} and resulting finally in 
the limit $\Lambda_{21} \to 0$. Clearly, Eq.~(\ref{15}) admits a countable set of roots 
$\{ \sigma_n \}_{n=1}^\infty$ or 
$\{ \lambda_n \}_{n=1}^\infty$ called resonances. Outside the resonances $\Lambda_{21} \to \infty$. 
If we consider the first repeated limit (\ref{7}) illustrated in Fig.~1 by path 1, the same cancellation as $l \to 0$ holds in the first square brackets, but now 
$\rho$ is non-zero and no cancellation of singularities occurs in the second  brackets, so that in this case always
$\Lambda_{21} \to \infty$. 

Consider now other paths in Fig.~1, like 3, 4, 5, which can be parametrized by the dependence $\rho =c \, l^\tau$ with any positive 
constants $c$ and $\tau$. In the $l \to 0$ limit the
cancellation of divergences ${\cal O}(l^{-1})$ again can be accomplished in the first square brackets of $\Lambda_{21}$, but now we must take into account the singularities  ${\cal O}(l^{\tau-2})$ appearing in the second brackets. Only in the case $\tau =1$ illustrated by path 3 these singularities can be cancelled out with those in the first brackets. As a result, the equation for resonances reads
\begin{equation}
{\tanh\sigma \over 1 +c\sigma \tanh\sigma} =\tan\sigma 
\label{16}
\end{equation}	
which in the particular case $c=0$ (i.e., $\rho =0$) coincides with
Eq.~(\ref{15}). Similarly, Eq.~(\ref{16}) also admits a countable set of solutions (resonances) $\{ \sigma_n \}_{n=1}^\infty$ or 
$\{ \lambda_n \}_{n=1}^\infty$. Clearly, no cancellation occurs for 
$\tau \in (0,2)\setminus \{ 1 \}$. This means that potential (\ref{2}) is opaque when the squeezing of the width $l$ occurs faster than the distance $\rho $ tends to zero [if $\tau \in (0,1) $, see path 5 in 
Fig.~1]. However, even in the case when the distance $\rho$ is squeezing faster than the width $l$, there are no resonances if $\tau \in (1,2)$. For the existence of resonances the squeezing of $\rho$ has to be more rapid compared with squeezing $l$ and this happens if $\tau \ge 2$. The exceptional case 
 $\tau =1$ with the resonances described by Eq.~(\ref{16}) falls under a general profile $v(\xi)$ of regularization (\ref{3}). 

As regards for $\tau \ge 2$, like path 4 in Fig.~1, the term 
${\cal O}(l^{\tau-2})$ in $\Lambda_{21}$ is no singular anymore. 
Its $l \to 0$ limit is zero for $\tau > 2$ and a non-zero constant at $\tau =2$. More precisely, this constant $g=g(\lambda) \doteq \lim_{l \to 0}\Lambda_{21} $  calculated at each resonance value $\sigma_n$, $n \in \N$, takes the following discrete values $g_n \doteq g(\lambda_n)$: 
\begin{equation}
  g_n = 
-c\sigma_n^2 \sinh\sigma_n \sin\sigma_n = { (-1)^{n+1} c \sigma_n^2
\sinh^2 \! \sigma_n \over \sqrt{\cosh(2\sigma_n)}}\, . 
\label{17}
\end{equation}
For this case the cancellation in the first square brackets of 
$\Lambda_{21}$ still takes place, so that the equation for resonances (\ref{15}) also holds for all $\tau \ge 2$.

Concerning the other matrix elements $\Lambda_{11}$ and 
$\Lambda_{22}$ as well as $\Lambda_{12}$, nowhere there exists a cancellation of divergences. The term $\Lambda_{12}$ has no singularities and therefore its zero-range limit is always zero, whereas  the limits of $\Lambda_{11}$ and $\Lambda_{22}$ depend on a chosen path in Fig.~1. Thus, following path 1 or alongside the path
$\rho = c\, l^\tau$ with $\tau \in (0,1)$, like path 5, one can see that $\Lambda_{11} \to \infty$ and $\Lambda_{22} \to \infty$. If we follow paths 2, 3 or $\rho = c\, l^\tau$ with $\tau \in (1, \infty)$,
 like path 4, in the zero-range limit $\Lambda_{11}$ and $\Lambda_{22}$ are finite everywhere. Moreover, at the resonance sets 
$\{ \lambda_n \}_{n=1}^\infty$ we have $\Lambda_{11} \to \chi $ and $\Lambda_{22} \to \chi^{-1}$ with $\chi=\chi(\lambda)$ taking discrete values $\chi_n \doteq \chi(\lambda_n)$, $n \in \N$. 
Thus, for the second repeated limit (\ref{7}) as well as for any path 
$\rho =c \, l^\tau$ with $\tau \ge 2$ we have 
\begin{equation}
\chi_n ={ \cosh\sigma_n \over \cos\sigma_n } = 
{ \sinh\sigma_n \over \sin\sigma_n }= (-1)^n 
\sqrt{\cosh(2\sigma_n) } \, .
\label{18}
\end{equation} 
Similarly, in the case of the path $\rho =c l$ we obtain the resonance values
\begin{eqnarray}
\chi_n &=&{ \cosh\sigma_n + c\sigma_n \sinh\sigma_n 
\over \cos\sigma_n } = 
{ \sinh\sigma_n \over \sin\sigma_n } \nonumber \\
&=& (-1)^n \sqrt{(\cosh\sigma_n +  c\sigma_n \sinh\sigma_n )^2 +
\sinh^2\! \sigma_n} \, .
\label{19}
\end{eqnarray} 

Thus, for any regularizing sequence $\Delta'_{l \rho}(x)$ which results in the existence of a corresponding resonance set
$\{ \lambda_n \}_{n=1}^\infty $ 
 the zero-range limit of Eqs.~(\ref{10}) and (\ref{11}) at these resonances becomes 
\begin{eqnarray}
\left( \begin{array}{cc} \psi(+0)   \\
\psi'(+0) \end{array} \right) = \Lambda \left(
\begin{array}{cc} \psi(-0)   \\
\psi'(-0)   \end{array} \right) , ~~~\Lambda = \left(
\begin{array}{cc} \chi~~~0~~ \\
g~~\chi^{-1}\! \!\end{array} \right) .
\label{20}
\end{eqnarray}
Here $\chi $ takes the discrete values  $\chi_n \neq 1 $,
 $n \in \N$, given by Eqs.~(\ref{18}) or (\ref{19}), while $g =0$ 
except for the case with $\rho = c \, l^2$ where the discrete values
 $g_n $, $n \in \N$, are calculated according to Eq.~(\ref{17}). Therefore in a resonance case one can define one of two mappings: either $\Delta'_{l \rho} (x) 
 \longmapsto \{ \chi_n \}_{n=1}^\infty $ with $g=0$ or 
$\Delta'_{l \rho} (x) 
 \longmapsto \{ \chi_n, g_n \}_{n=1}^\infty $, so that running in a general case over different regularizing sequences $\Delta'_\varepsilon (x)$, in the 
$\varepsilon \to 0$ limit one can obtain a whole family of matrix sequences $\{ \Lambda_n \doteq \Lambda(\lambda_n)\}_{n=1}^\infty$. Clearly,
boundary conditions (\ref{20}) are invariant under the transformation 
$ \psi(\pm 0) \to \chi \psi(\mp 0)$ and $ \psi'(\pm 0) \to \chi^{-1} \psi'(\mp 0)$. They form a subfamily of the whole family of non-separated connection matrices \cite{adk} 
\begin{eqnarray}
\Lambda = {\rm e}^{{\rm i} \vartheta}\left(
\begin{array}{cc} \lambda_{11}~~ \lambda_{12} \\
\lambda_{21} ~~ \lambda_{22} \end{array} \right)
\label{21}
\end{eqnarray}
with real parameters $\vartheta \in [0,\pi)$ and  
$\lambda_{ij} \in \R$, $i,j =1,2$, fulfilling the condition  
$ \lambda_{11}\lambda_{22} - \lambda_{12}\lambda_{21}  = 1 $.

The zero-range limit of 
the reflection and transmission coefficients $R$ and $T$ can also be 
 given in terms of $\chi$ and $g$. Indeed, as follows from 
Eqs.~(\ref{14}), we obtain
\begin{equation}
R = { \chi^{-1} - \chi - {\rm i}g/k 
\over  \chi^{-1} + \chi + {\rm i}g/k } ~~~
\mbox{and}~~~
T = { 2 \over  \chi^{-1} + \chi + {\rm i}g/k } \,  .
\label{22}
\end{equation}
Particularly, in the case of the resonances described by 
Eq.~(\ref{15}) we obtain $R = -\tanh^2 \! \sigma_n$ and 
$T = (-1)^n \sqrt{1 -\tanh^4 \! \sigma_n}\, $.
Outside the resonances we have $g = \infty$, 
 so that $R = -1$ and $T = 0$. 
The case $\chi=1$ with $g \neq 0$ corresponds to the pure $\delta$-interaction,
while the case with $\chi \neq 1$ to potential (\ref{2}). 
The case when both $\chi \neq 1$ and $g \neq 0$ can be treated as the 
$\delta'$-interaction accompanied by an effective $\delta$-potential.
Therefore one can expect the existence of a non-trivial bound state
with energy $E \doteq -\kappa^2$ if $g \neq 0$.  Indeed,
looking for negative-energy solutions of Eq.~(\ref{1}) in the form
\begin{equation}
\psi (x) =  \left\{ \begin{array}{ll}
   A \, {\rm e}^{\kappa x}
 &  \mbox{for}~~ - \infty < x < 0,  \\
B \, {\rm e}^{-\kappa x}  & 
  \mbox{for}~~~~~ ~ 0 < x <  \infty   ,
\end{array} \right.
\label{23}
\end{equation} 
one can write the matrix equation
\begin{eqnarray}
 \left( \begin{array}{cc} B \\
 -\kappa B \end{array} \right) 
 = \left( \begin{array}{cc} \!\! \!\! \! \chi~~ 0 \\
g ~~ \chi^{-1} \end{array} \right) \left(
\begin{array}{cc} A \\
 \kappa A  \end{array} \right) .
\label{24}
\end{eqnarray}
The compatibility of solutions for this equation gives
the equation for $\kappa$ from which, when using Eqs.~(\ref{17}) and (\ref{18}), we immediately obtain for each $\lambda_n$,
 $n \in \N$,  one bound state with  
\begin{equation}
\kappa_n = - \, {g_n \over \chi_n +\chi^{-1}_n } = {c \over 2} \sigma_n^2 \tanh^2 \! \sigma_n \, .
\label{25}
\end{equation}

\section{The connection matrix obtained through a generalized distribution theory}

Now, we would like to find a possible interpretation of the matrices $\Lambda_n$, $n \in \N$, in terms of distributions, similarly to Albeverio et al.  \cite{adk}, where
 potential (\ref{2}) has been considered as a particular example in the general theory of self-adjoint extensions for point interactions. One could follow  Kurasov's  extension of the distribution theory to the space of test functions discontinuous at the origin \cite{ku}. This extension is based on the suggestion to
define the distributions $\delta^{(n)}(x)$ on the space of test
functions $\psi(x)$ discontinuous at $x=0$ through the averaged formula
\begin{equation}
\langle \delta^{(n)} | \psi \rangle
 = (-1)^n { \psi^{(n)}(-0) + \psi^{(n)}(+0) \over 2} \, 
\label{26}
\end{equation}
with the equal ($1/2$) weights at the left and right limits of the function $\psi(x)$ and its derivatives $\psi^{(n)}(x)$ at the origin \cite{gr,kse}. Particularly, for $n=1$ the well defined product
\begin{equation}
\delta'(x) \psi(x) = \psi(0) \delta'(x) -\psi'(0)\delta(x)
\label{28}
\end{equation}
 for any continuous function $\psi(x)$
and its continuous derivative were supposed  to be generalized as 
\begin{equation} 
\delta'(x) \psi(x) = { \psi(-0) + \psi(+0) \over 2}
 \delta'(x) - { \psi'(-0) + \psi'(+0) \over 2 } \delta(x) .
\label{30}
\end{equation}
As a result, the boundary conditions for potential (\ref{2}) in the
form of the diagonal matrix \cite{adk,gr,kse}
\begin{eqnarray}  
\Lambda = \left(
\begin{array}{cc}  \!\!\! \! {\cal A} ~~~0~ \\ 
0~ ~{\cal A}^{-1}
\end{array} \right), ~~~{\cal A} = {2 + \lambda \over 2 -\lambda} \, ,
\label{31}
\end{eqnarray}
have been established and afterwards
used in many studies (see, e.g., \cite{cnp,kse,ni}).
When additionally in potential (\ref{2}) the term 
$\gamma \delta(x)$ is included, matrix (\ref{31}) is modified to 
\cite{adk,gnn}
\begin{eqnarray}  
\Lambda = \left(
\begin{array}{cc} ~~~{2 + \lambda 
\over 2 -\lambda} ~~~~~~~0~~~~\\
{\gamma \over 1 - \lambda^2/4}  ~~ ~~ {2 - \lambda  \over 2 +
\lambda}\end{array} \right) .
\label{32}
\end{eqnarray}
Note that throughout the present paper we are dealing only with the pure $\delta'(x)$ potential given by Eq.~(\ref{2}).

The common feature of the results obtained by the regularization
procedure \cite{c-g,zci,gm} and those obtained within the theory of self-adjoint extensions \cite{adk} is only the form of matrices 
(\ref{20}) with $g=0$ and (\ref{31}). The discrete values of $\chi$ calculated according to Eqs.~(\ref{18}) or (\ref{19}) cannot be superposed on the graph of the function ${\cal A}(\lambda)$ defined in 
(\ref{31}) because of obvious quantitative difference. Consequently, it is impossible to achieve any compatibility of the results obtained by regularization and matrix (\ref{31}). In this regard, one can think that distributions are not the mathematically  rigorous concept for zero-range interactions. In favour of this concept is also the fact that despite \cite{ku}, there exists no appropriate distribution theory for discontinuous test functions. Nevertheless, below we shall 
start with a general expression for the potential term and single out those values of parameters which give a non-zero transmission using however only the standard definition of distributions for the 
$C^\infty$ test functions.

Thus, as demonstrated above for the resonant case, the cancellation of singularities in the zero-range limit occurs in a different way depending on chosen regularizing sequence $\Delta'_\varepsilon(x)$. This means that in the regularization scheme one or more hidden parameters should be present which control the process of realization of the zero-range limit, despite $\delta'(x)$ itself does not contain any parameters. On the other hand, the potential term in 
Eq.~(\ref{1}) contains the ambiguous product 
$\delta'(x) \psi(x)$ [$\psi(x)$ is discontinuous at $x=0$] and  therefore these parameters can be involved into the definition of this product using, like Eq.~(\ref{30}),  the classical $\delta(x)$ and 
$\delta'(x)$ distributions. In other words, instead of the function ${\cal A}(\lambda)$ defined in (\ref{31}) we should incorporate a whole family of functions 
${\cal A}(\alpha ; \lambda)$ depending, at least, on one parameter, say $\alpha$. Then for each resonance value 
$\lambda = \lambda_n$, $n \in\N$, given by (\ref{15}) or 
(\ref{16}) one can try to find such a value $\alpha = \alpha_n$ that satisfies the equation
$\chi_n = {\cal A}(\alpha, \lambda_n)$ with $\chi_n$ given by 
(\ref{18}) or (\ref{19}). If this case happens for any $n \in \N$, one could claim then that the cancellation of singularities occurs exactly at that value $\alpha = \alpha_n$ which corresponds to a given 
$\chi_n$. In this regard, it does not matter that  
$\delta(x)$ is even and $\delta'(x)$ odd.  The only requirement is that the product  
$\delta'(x) \psi(x)$ has to be a linear combination of the classical  
$\delta(x)$ and $\delta'(x)$ distributions. Therefore instead of 
Eq.~(\ref{30}), one can suggest to define the ambiguous product 
$\delta'(x) \psi(x)$ as follows
\begin{eqnarray} 
 \delta'(x)\psi(x) &=& \left[(1-\alpha ) \psi(-0) 
+\alpha \psi(+0) \right] \delta'(x) \nonumber \\
&-& \left[ \alpha \psi'(-0) + (1- \alpha) \psi'(+0)
\right]\delta(x) \nonumber \\
&+& \beta \left[ \psi(+0) -\psi(-0)\right] \delta(x)  
\label{33}
\end{eqnarray}
with arbitrary coefficients $\alpha$ and $\beta$. This expression makes sense since it is a linear combination of the distributions 
$\delta(x)$ and $\delta'(x)$ defined on the $C^\infty$ test functions. The particular case with 
$\alpha =1/2$ and $\beta =0$ coincides with definition 
(\ref{30}). The presence of the last term here with the coefficient 
$\beta$ is motivated by the necessity of having the boundary conditions with $g \neq 0$ in Eq.~(\ref{20}). Clearly, for continuous functions $\psi(x)$ and $\psi'(x)$ Eq.~(\ref{33}) reduces to well defined relation (\ref{28}). Note that formula (\ref{33}) is postulated, similarly to formula 
(\ref{30}) postulated earlier by Griffiths \cite{gr} and using 
afterwards by other authors (see, e.g., \cite{ku,kse,ni,gnn}) as a key point for their studies. 

In order to obtain the connection between the two-sided boundary conditions described by matrix (\ref{20}), one could follow Griffiths \cite{gr}, i.e., integrate Eq.~(\ref{1}) from $-\epsilon$ to $\epsilon$ and then accomplish the $\epsilon \to 0$ limit.
 Instead, here we prefer the approach of Gadella et al. 
\cite{gnn}, when one can  control the cancellation of singularities in the process of calculations. 
Therefore, similarly to \cite{gnn}, we represent the wave function 
$\psi(x)$ in the form
\begin{equation}
\psi(x) =\psi(-0){\rm e}^{ -{\rm i}kx} \Theta(-x)
+ \psi(+0){\rm e}^{ {\rm i}kx} \Theta(x) ,
\label{34}
\end{equation}
where  $\Theta(x)$ is the unit step function.  Representation 
(\ref{34}) describes the waves propagating to the left and to the right from the origin and its second distributional derivative is
\begin{eqnarray}
\psi''(x) &= &- k^2 \psi(x) + 2{\rm i} k \left[ \psi(-0)
{\rm e}^{-{\rm i}kx } + \psi(+0){\rm e}^{{\rm i}kx } \right] 
\delta(x) \nonumber \\
&+& \left[ \psi(+0){\rm e}^{{\rm i}kx } -\psi(-0)
{\rm e}^{-{\rm i}kx }\right] \delta'(x).
\label{35}
\end{eqnarray}
Here in the square brackets we have the $C^\infty$ functions and therefore one can exploit Eq.~(\ref{28}) together with 
the relation $\delta(x) \psi(x) = \psi(0) \delta(x)$.  Using finally the relations $- {\rm i}k\psi(-0) = \psi'(-0)$ and 
$ {\rm i}k\psi(+0) = \psi'(+0)$ obtained directly from 
Eq.~(\ref{34}), we immediately find
\begin{equation}
\psi''(x) = - \, k^2 \psi(x) + \left[ \psi'(+0) -\psi'(-0)\right] 
\delta(x) + \left[ \psi(+0) -\psi(-0)\right] \delta'(x) .
\label{36}
\end{equation}
Next, we insert the right-hand-sides of Eqs.~(\ref{33}) and 
(\ref{36}) into Eq.~(\ref{1}). The latter equation will be satisfied if the coefficients  at $\delta(x)$ and $\delta'(x)$ cancel out. As a result, we obtain the boundary conditions 
\begin{eqnarray}
&& (1 - \alpha \lambda) \, \psi(+0) =[ 1 + (1-\alpha) \lambda ]
\, \psi(-0) , \nonumber \\
&& [ 1 + (1-\alpha) \lambda ] \, \psi'(+0) =(1 - \alpha \lambda) 
\, \psi'(-0) + \beta \lambda \left[  \psi(+0) - \psi(-0)\right]
\label{37}
\end{eqnarray}
 which can be rewritten in the form of Eq.~(\ref{20}) with the connection matrix 
\begin{equation}
\Lambda = \left(
\begin{array}{cc}  \!\!\! \! {\cal A} ~~~0~ \\ 
{\cal B}~ ~{\cal A}^{-1} \end{array} \right) 
\label{37a}
\end{equation}
where
\begin{equation}
{\cal A}(\alpha; \lambda) = {1 +( 1-\alpha) \lambda \over 1-\alpha \lambda} ~~\mbox{and}~~
 {\cal B}(\alpha, \beta; \lambda) = {\beta \lambda^2 \over 
(1 -\alpha \lambda) [1+ (1-\alpha) \lambda] } \,.
\label{38}
\end{equation}
In the particular case $\alpha = 1/2$ and $\beta =0$ matrix 
(\ref{37a}) with elements (\ref{38}) coincides with matrix (\ref{31}).
Having the free parameters $\alpha$ and $\beta$ in expressions
(\ref{38}), for any resonance set 
$\{ \lambda_n \}_{n=1}^\infty$ given, for instance, by 
(\ref{15}) or (\ref{16}) one can solve each of the compatibility equations $\chi_n = {\cal A}(\alpha ; \lambda_n)$ and $g_n = {\cal B}(\alpha, \beta ; \lambda_n)$ with respect to these parameters. As a result, one finds 
\begin{equation}
\alpha_n = {1 \over \lambda_n} + {1 \over 1 - \chi_n}~ ~~
\mbox{and}~~~
\beta_n = { \chi_n g_n \over (1 -\chi_n )^2} \, 
\label{40}
\end{equation}
as functions of $\lambda_n$, $\chi_n$ and  $g_n$,
 $n \in \N$.  Consequently, for any resonance value $\lambda_n$ there exists the one-to-one correspondence between 
$ \chi_n \in \R \setminus \{1\} $ calculated within the regularization approach  and  
$\alpha_n$ in (\ref{33})  if $g =0$. The similar correspondence takes place between  the pairs $\{ \chi_n, g_n \}$ and 
$\{ \alpha_n , \beta_n \}$ if $g \neq 0$.

Finally, it should be mentioned that the same boundary conditions, given by matrix (\ref{37a}) with elements (\ref{38}) can be obtained if we start with the presentation of the wave function $\psi(x)$ for negative-energy solutions (bound states) using again the 
Gadella-Negro-Nieto approach \cite{gnn}, i.e.,
\begin{equation}
\psi(x) =\psi(-0){\rm e}^{ \kappa x} \Theta(-x)
+ \psi(+0){\rm e}^{ -\kappa x} \Theta(x) , ~~ \kappa \doteq 
\sqrt{-E} \, .
\label{39}
\end{equation} 
Indeed, differentiating this expression twice as above and using the same formula (\ref{33}), one obtains the connection matrix $\Lambda$ of the same form (\ref{37a}) with the elements ${\cal A}$ and 
${\cal B}$ given by expressions (\ref{38}) as expected.

\section{Concluding remarks}

It has been demonstrated on simple rectangular model (\ref{6})  regularizing singular potential (\ref{2}) with two squeezing parameters $l$ and $\rho$ that the reflection-transmission results appear to be quite different depending on a chosen regularizing sequence $\Delta'_{l \rho}(x)$ shown schematically in Fig.~1 by a path. In dependence what squeezing ($l$ or 
$\rho$) is faster compared with the other one, we have observed either the full reflection or a partial resonant tunnelling at a countable set $\{ \lambda_n \}_{n=1}^\infty$ in the 
$\lambda$-space. In its turn, the structure of a resonance set also depends on a chosen path. The existence of resonance sets has been shown to be a result of cancellation in the zero-range limit of divergences emerging from the kinetic energy and potential terms. The boundary conditions in the case of resonant tunnelling are given by the connection matrix $\Lambda$ of form (\ref{20}) where $\chi$ and $g$ take finite values only at the resonances 
[see Eqs.~(\ref{17})-(\ref{19})].  

On the other hand, the connection matrix $\Lambda$ given by 
Eq.~(\ref{31}) and realized by distributions conflicts with that obtained by regularization [see Eqs.~(\ref{17})-(\ref{20})]. As a result, it appears that the way through regularizing sequences and the way through singular distributions lead to different results. Because of this discrepancy, one could conclude that the singularities cancelled out under regularization are somewhat different from singular distributions. This might be an interesting problem for further studies, nevertheless, in the present paper we would like to enforce a relationship between the regularization approach and the distribution theory removing the above-mentioned discrepancy. To this end we have enlarged the family of self-adjoint extensions involving into the ambiguous product $\delta'(x)\psi(x)$ two free parameters 
$\alpha$ and 
$\beta$. The boundary conditions obtained in this way through connection matrix (\ref{37a}) with elements (\ref{38}) appear to be of a general form. In particular, they recover the results obtained through the regularization procedure with discrete values $\alpha =\alpha_n$ and $\beta =\beta_n$ calculated according to 
Eqs.~(\ref{40}). 
Despite the $\delta'(x)$ distribution does not contain any free parameters, in the renormalization scheme they are present as hidden parameters. These parameters control the process of cancellation of singularities resulting in specific values 
$\lambda_n$, $\chi_n$ and $g_n$ presented by 
Eqs.~(\ref{15})-(\ref{19}). Therefore the hidden parameter values 
$\alpha_n$ and $\beta_n$ given by Eqs.~(\ref{40}) correspond exactly to that regularizing sequence which leads to a chosen triple 
$ \{ \lambda_n, \chi_n , g_n \}$. This relationship seems to be the main motivation for modifying the theory of self-adjoint extensions through generalized postulate (\ref{33}) with the free parameters 
$\alpha$ and $\beta$.  

Finally, note that the estimates of $\lambda_n$'s and 
$\chi_n$'s performed in Eqs.~(\ref{15}), (\ref{16}), (\ref{18}) and 
(\ref{19}) give for the first equation (\ref{40}) the inequalities 
$0 < \alpha_n <1$ for all $n \in \N$. Only for a non-resonant case with $\chi = 1+\lambda$ to be published elsewhere we have $\alpha =0$.

\bigskip

\noindent
{\bf Acknowledgments}
\bigskip

\noindent
We are very grateful to anonymous Referee for valuable suggestions, comments and critical remarks,
resulting in an essential revision and improvement of the final version of this paper. Stimulating and helpful discussions with 
Yu.D. Golovaty,
P.L. Christiansen, Y.B. Gaididei and S.V. Iermakova are greatly 
acknowledged.

\end{document}